# Simulation of the Efficiency of a-SiC:H/a-Si:H Tandem Multilayer Solar Cells

Khikmat Kh. Muminov, *)Ashrafalsadat S. Mirkamali

*S.U.Umarov Physical-Technical Institute, Academy of Sciences of the Republic of Tajikistan, 299/1 Aini Ave, Dushanbe 734062, Tajikistan*
*e-mail: khikmat@inbox.ru*
*) *Permanent address: Department of Science and Engineering, Behshahr Branch, Islamic Azad University, Behshahr, Iran*
*e-mail: ash.mirkamali@gmail.com*

In this paper we carried out theoretical study of the general issues related to the efficiency of SiC:H/a-Si:H single- and multi-junction tandem solar cells. Implementation of numerical simulations by the use of AMPS-1D program of one-dimensional analysis of microelectronic and photonic structures for the analysis of hydrogenated silicon solar cells allowed us to formulate the optimal design of new kind of multi-junction tandem solar cells, providing its most efficient operation. The numerical analysis of SiC:H/a-Si:H single-junction solar cell whith doped i-layer used as the intermediate absorbing layer (a -Si: H) placed between layers of p-type (a-SiC: H) and n-type (a-Si: H) has been conducted. It has been established that after optimizing the solar cell parameters its highest efficiency of 19.62% is achieved at 500 nm thickness of i-layer. The optimization of the newly developed multi-junction structure of a-SiC:H/a-Si:H tandem solar cell has been conducted. It has been shown numerically that its highest efficiency of 22.6% is achieved at the thickness of 270 nm of intermediate i-layer.

## 1. Introduction

Nowaday, the role and significance of renewable energy is becoming increasingly important due to the fact that the main sources of energy (i.e. oil, coal, uranium, etc.) are limited and consequences of their use to the natural balance on our planet are becoming increasingly apparent. Plants are not able to absorb the huge amount of carbon dioxide that is emitted into the Earth's atmosphere from the burning of fossil fuels [1, 2]. As a result the solar energy reaching the earth's surface acquired the increasing significance, which we can use in order to convert it into clean electricity.

Solar cells are photovoltaic (or photoelectric, further PV) devices, which convert electromagnetic radiation from the sun (i.e., light, including infrared, visible and ultraviolet) into electricity. To use them in practical applications, photovoltaic devices should meet numerous requirements. The first requirement is the conversion efficiency of solar energy into electricity. Second, the material used should be inexpensive, available in large quantities and nontoxic. Third, the method of producing the device should be inexpensive, energy-efficient, fast, simple and environmentally safe. Fourthly, productivity of a solar cell must be stable for a long period of time.

Currently there is a significant need to develop low cost solar cells with high efficiency and stability. In thin-film technology there is a common problem associated with the conversion efficiency due to the low quality of materials; there a rapid recombination of photogenerated electrons and holes takes place, whereby the free charge carriers have a short diffusion length and, hence, losses in energy conversion occur. If solar cells are based on nanoscale heterojunctions, each photogenerated charge carrier must pass a much shorter distance and the problem of recombination can be mainly overcome [3].

Multijunction solar cells are created from multiple p-n junctions of different semiconductor materials with different band gap, so that they have the ability to absorb most of the energy of the solar spectrum. Single-junction solar cells during the absorption of photons of solar spectrum have some limitations, such as the absorption of photons of the solar spectrum may occur only in a range from 300 nm to 2500 nm. In fact, they do not have the ability to absorb photons with wavelengths longer than 1100 nm, which contains more than 20% of the radiation at the AM 1.5 standard spectra [5]. The photons with a short wavelength in the ultraviolet region due to thermal effects can not be effectively converted hence if multilayer cells based on different materials having different band gaps, it is possible to increase the effective absorption of photons, and thus increase the efficiency of solar cell.

For the numerical simulation of a number of highly sophisticated semiconductor devices a program of one-dimensional modeling and analysis of microelectronic and photonic structure AMPS-1D is used. This software can be used to analyze transport phenomena in a wide variety of structures and devices that may comprise a combination of crystalline, polycrystalline or amorphous semiconductor layers. Program AMPS-1D for one-dimensional device simulation has been developed by Professor S. Fonash and colleagues (University of Pennsylvania, USA).

In this paper, we study multijunction tandem solar cells that are based on inorganic semiconductor materials solar cells of the second generation, in order to obtain maximum efficiency. To improve the efficiency of the solar cell based on hydrogenated amorphous silicon a-Si:H, the main attention should be paid mainly to obtaining a high short-circuit current $J_{sc}$ by developing light traps [6], as well as a higher open-circuit voltage $V_{oc}$, by applying new methods of device designing using different alloys of a-Si.

One of the new solar cell structures used hydrogenated microcrystalline silicon (μc-Si: H) as the active layer of the solar cell, it was first proposed by Swiss research group in 1994 [7]. They had received the initial efficiency of 4.6% for the p-i-n-singlejunction solar cells based on μc-Si: H, practically without light-induced degradation of cell. However, they have also developed the so-called

"micromorphic" concept of the a-Si/μc-Si tandem solar cell, and reached 12% stabilized efficiency [8].

The light trap in this device is an intermediate reflector in the micromorphic solar cell; it reduces the light-induced degradation by approximately 1.64% [9]. Several research groups attempted to increase the open circuit voltage $V_{oc}$, using, for instance, the semiconductor materials with wide band gaps (such as a-SiC: H, μc-SiO: H, nc-Si:H , and so on) as window layers in amorphous silicon a-Si solar cells, also they used intermediate buffer layers in the p/i and i/n junctions [10, 11]. For example, Saito et al [12] in 2005, proposed a triple junction a-Si:H/μc-Si:H/μc-Si:H solar cell with an initial efficiency of 13.1% and light-induced degradation of about 7.2%.

## 2. Thin Film a-Si:H Solar Cells

Crystalline (and polycrystalline) silicon has an indirect band gap, which results in low optical absorption coefficient. The reason why the silicon material is so dominant is the fact that silicon is a semiconductor with a well-balanced set of electronic, chemical and physical properties. It is important that the technology for silicon producing is well designed for electronic devices. In order to create high-performance solar cells, it is important to be able to control and limit the diffusion of atoms in the intermediate i-layer in the solar cell production process [13].

Hydrogenated amorphous silicon (a-Si:H) solar cells are fairly well studied, and their production is developed as a cheaper alternative as compared with the better single crystal c-Si devices. Although solar cells made of this material have much less efficiency (less than 10% [14]) as compared to c-Si, but the films are used with the same thickness as those used in thin-film photovoltaics, due to the much higher absorption coefficient of a-Si, in comparison with a transparent semiconductor. However, these solar cells are susceptible to light-induced degradation, which eventually leads to a drop in efficiency up to 6-7% and stabilizes in this range [15].

A new kind of solar cell that attracts the attention of researchers, are solar cells based on “heterojunction with internal thin layer” [16].

Schematic sketch of a standard a-Si:H solar cell is shown in the Fig. 1. Manufacturing of this solar cell usually begins on the stainless steel. Indium tin oxide (ITO) $In_2O_3$-$SnO_2$ layer is typically applied as a transparent conductive oxide (TCO). The thickness of the ITO layer is about 200 nm, and the thickness of TCO varies considering the conductivity and optical transparency. This cell uses p-type layers of a-SiC:H semiconductor deposited on n-type a-Si:H semiconductor in order to form a p-n-heterojunction, which is an inner layer between n-type layers of a-Si [17].

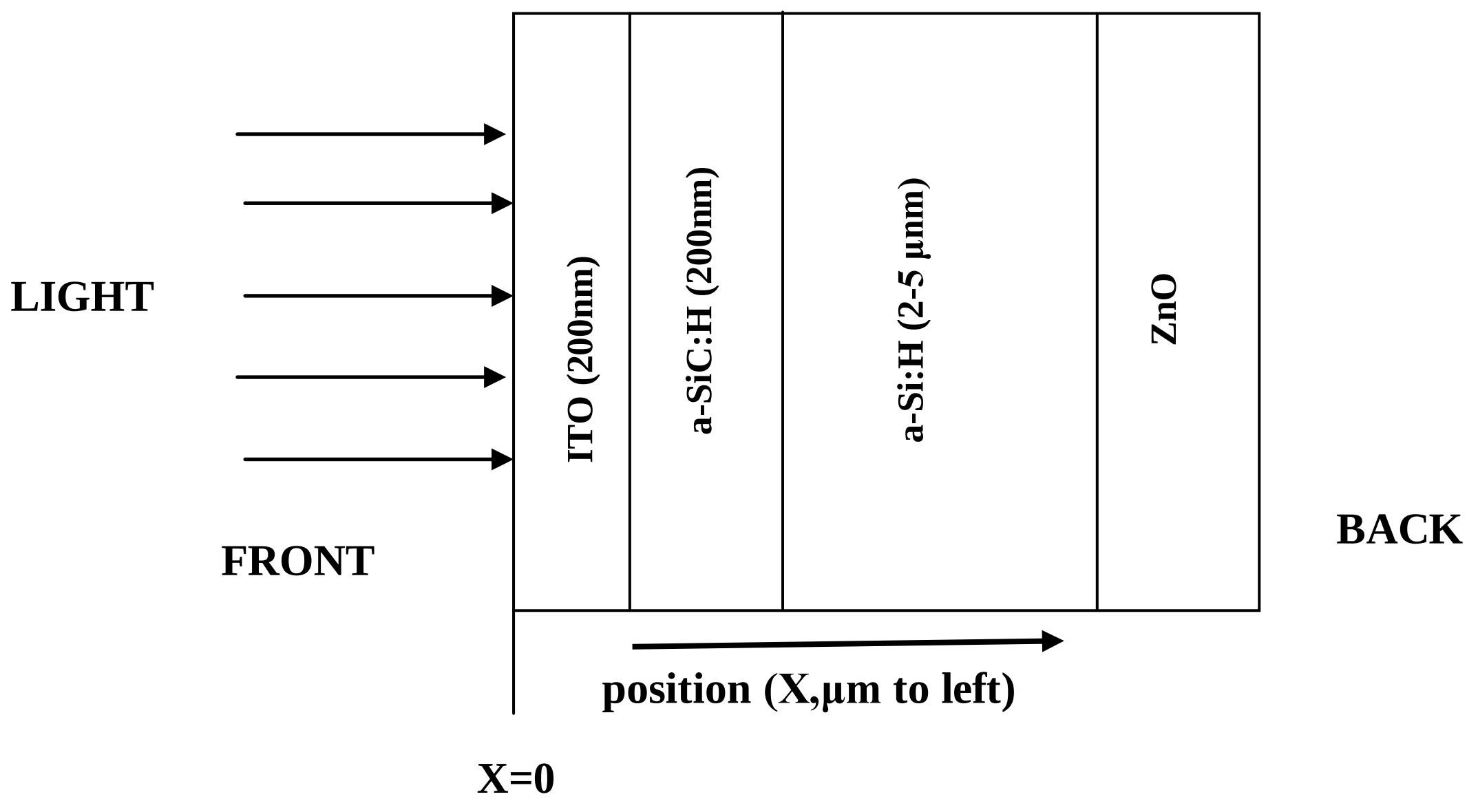

**Fig 1.** The typical structure of the hydrogenated amorphous silicon (a-SiC:H/a-Si:H) solar cell

One of the promising optoelectronic materials for applications in solar cells is hydrogenated amorphous silicon a-Si:H. Thin film a-Si:H has a high optical absorption coefficient (>$10^5$ $cm^{-1}$), a tunable bandgap and low temperature of deposition. The bandgap of a-Si:H can range from 1.6 to 1.8 eV. The band gap of 1.5 eV gives the highest conversion efficiency.

Progress in the use of hydrogenated amorphous silicon a-Si:H solar cells began with the invention of the first Schottky device with an efficiency of 2.4% by Carlson and Wronski [18]. The conversion efficiency of single-junction thin-film a-Si:H solar cells has been gradually improved from 2.4% [18] to 10.1% [19].

In carrying out the simulation of the device, to reduce losses due to reflection, the front surface is considered as a transparent contact made of a conductive oxide ($SnO_2$:F), and p-layer serves as a window through which the light enters the solar cell. Between the p-type (a-SiC: H) and n-type (a-Si: H) layers a doped layer is placed, which is used as an intermediate absorbing layer (a-Si: H). Photons absorbed in the i-layer generate an electron-hole pair. The electric field induced along the i-layer by the p- and n-layers leads to the drift of electrons to the n-layer and holes to the p-layer. In the simulation of the device due to the width of a-SiC:H layer bandgap, it is used as a front absorbing doped p-layer, in order to reduce the loss caused by absorption. To reduce material costs and increase efficiency of the solar cell it is necessary to optimize the thickness of the i-layer. For this purpose we perform numerical simulation use AMPS-1D software. The parameters of different materials of single-junction a-Si:H solar cells used for numerical simulation are shown in Table 1.

In order to find the optimal thickness and effectiveness in the course of numerical simulation we very the thickness of i-layer in ranges from 100 to 1300 nm. The maximum efficiency achieved was 19.62% at 500 nm for a thickness of i-layer, as shown in Figure 2. A similar result was also obtained in paper [21]. In this paper, the optimized thickness of the inner layer (500 nm) was significantly reduced compared with other studies of 700 nm [20], 600 nm [22], and 840 nm [23], respectively. Thus, we performed a numerical analysis of a single-junction a-Si:H solar cells, and found that after its optimizing the highest efficiency in 19.62% achieved at the intermediate i-layer thickness of 500 nm. The maximum experimental value of the similar solar cell efficiency is 10.1% (Kabir et al. [4]).

**Table 1.** Input data for simulation of a-Si:H/a-SiC:H single-junction solar cell.

| Material | Band gap (eV) | Conductivity type | Conduction Band | Valence Band | Electron Affinity (eV) | Electron Mobility (cm2 /v/s) | Hole Mobility (cm2 /v/s) | Free Carrier Concentration ($cm^{-3}$) | Relative Permittivity |
|---|---|---|---|---|---|---|---|---|---|
| **$SnO_2$:F** | 3.7 | P | $2.2*10^{18}$ | $1.8*10^{19}$ | 4.8 | 60.0 | 6.0 | $1.3*10^{19}$ | 9.0 |
| **a-SiC:H** | 1.9 | P | $2.5*10^{20}$ | $2.5*10^{20}$ | 4.0 | 20.0 | 2.0 | $3.0*10^{18}$ | 11.9 |
| **a-Si:H** | 1.75 | N | $2.5*10^{20}$ | $2.5*10^{20}$ | 3.8 | 20.0 | 2.0 | $8.0*10^{19}$ | 11.9 |
| **ZnO:B** | 3.3 | N | $2.2*10^{18}$ | $1.8*10^{19}$ | 4.5 | 33.0 | 8.0 | $8.0*10^{18}$ | 9 |

**Fig. 2.** Dependence of output parameters of single-junction a-Si:H/a-SiC:H solar cells on the thickness of absorbing layer

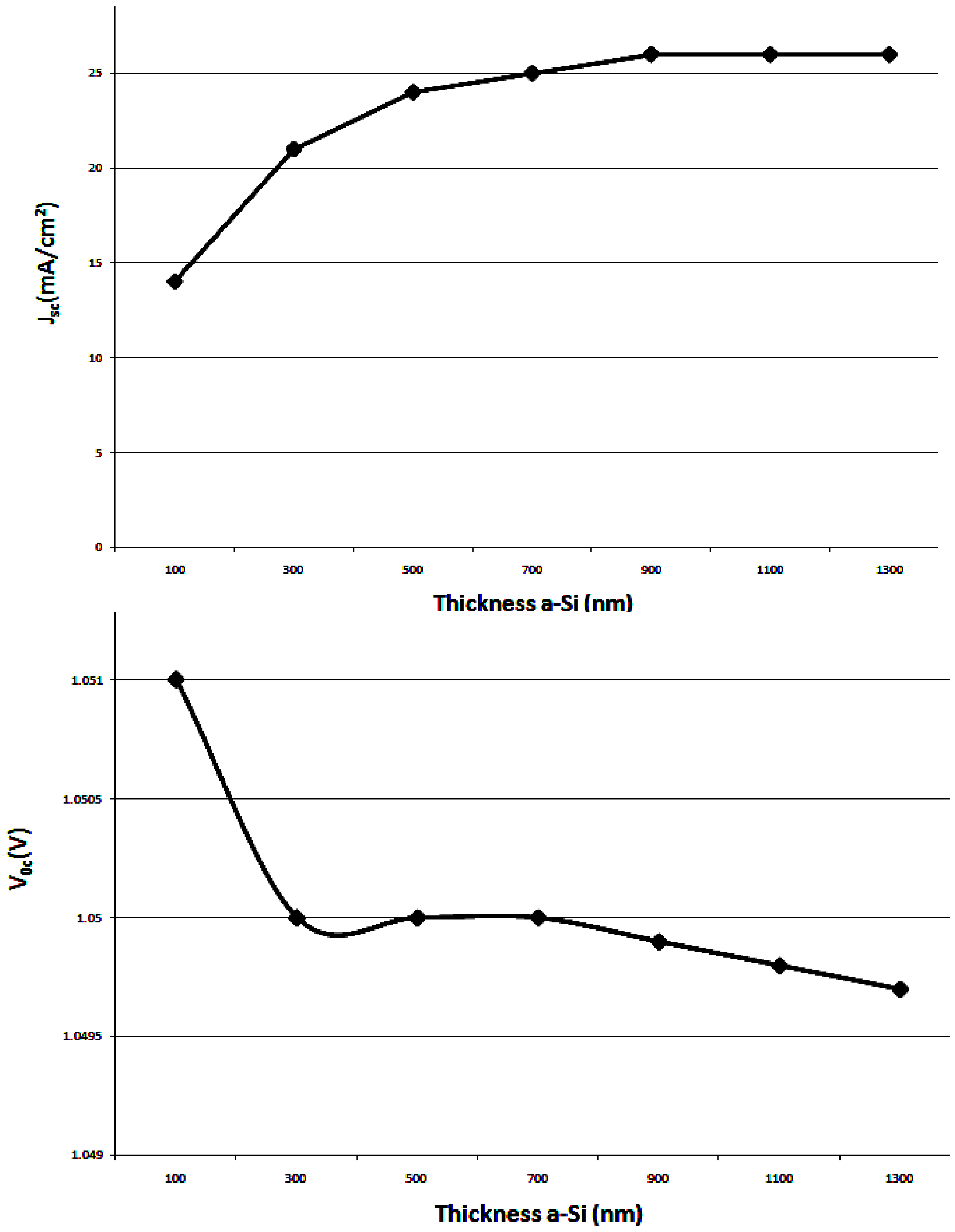

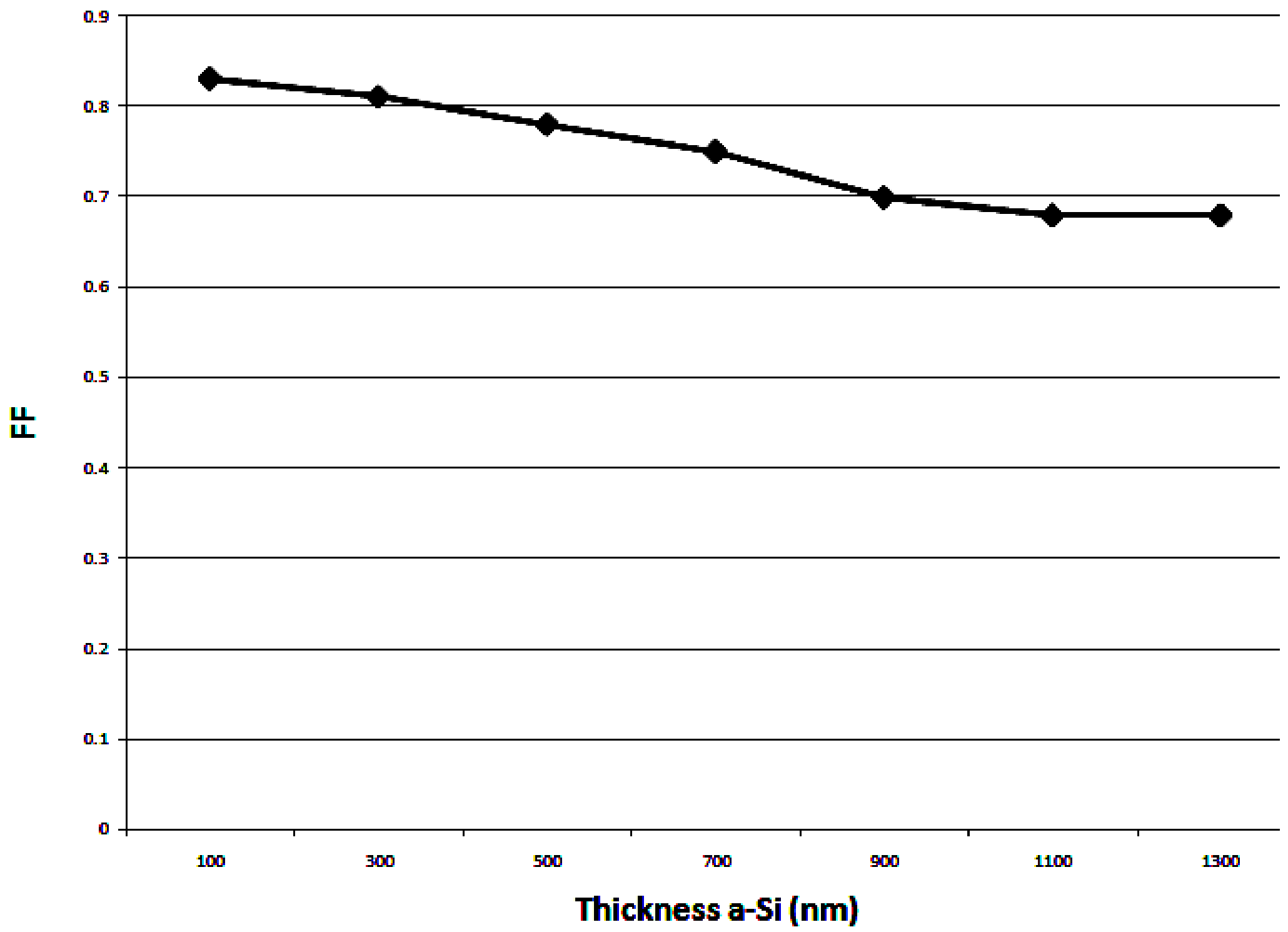

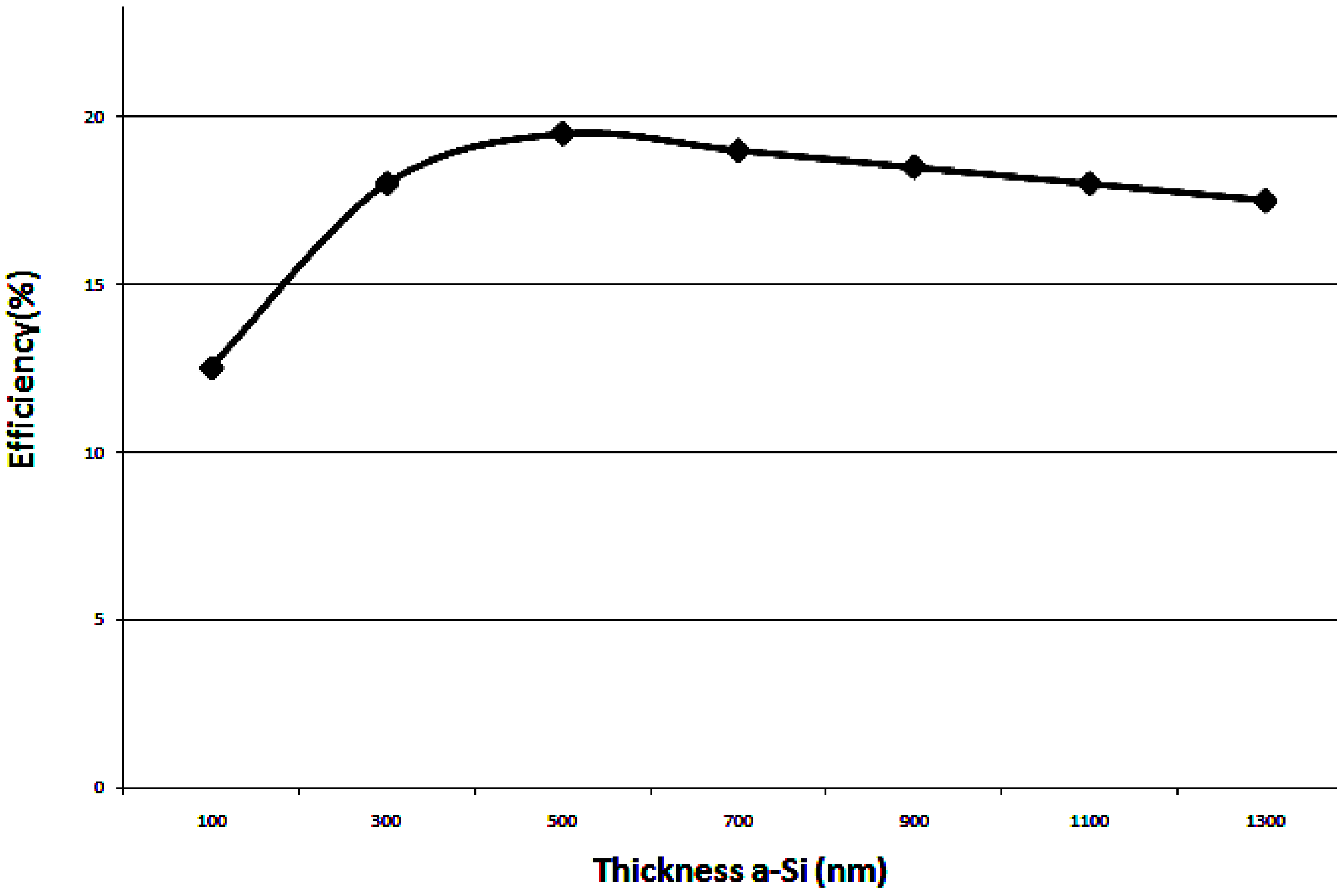

## 3. Tandem multijunction a-SiC:H/a-Si:H solar cells

Multijunction solar cells are created on the basis of multiple p-n-junctions of different semiconductor materials with different band gaps, which have the ability to absorb most of the energy of the solar spectrum. Single-junction solar cells during the absorption of photons of the solar spectrum have some limitations, such as the absorption of photons of the solar spectrum may occur only in a range from 300 nm to 2500 nm. In fact, they do not have the ability to absorb photons with wavelengths longer than 1100 nm, which contains more than 20% of the radiation at the standard AM 1.5 spectrum [5]. Photons with a short wavelength in the ultraviolet region due to thermal effects can not be effectively converted hence if multilayer cells based on different materials having different band gaps are used, it is possible to increase the effective absorption of photons, and thus increase the efficiency of solar cell.

Multi-junction solar cells on the basis of elements of III-V groups represent the new technology as compared with the single-junction solar cells that provide at the same conditions, nearly twice the efficiency [24], but the production costs of these cells are very high, and they are used only in special cases, for example in space technologies. Multijunction solar cells, compared with single-junction ones, possess two significant advantages:

a) Absorption of a larger number of incident photons.

The band gap of multi-junction solar cells (MSC) is much smaller than of the single-junction (SSC) ones. Consequently, the MSC absorb a larger number of photons having lower energy than the band gap of single-junction solar cell (SSC), and therefore, could not participate in the production of electron-hole pairs. Moreover, MSC absorb photons having high energy, which excess in SSC is lost as heat. Accordingly, MSC not only absorb a larger number of photons of the incident light, but also produces a large number of electron-hole pairs, compared with the SSC. However, in fact, less current is generated because each layer absorbs photons only of part of the solar spectrum, and therefore, more voltage is created.

b) Another advantage of MSC is low power losses due to the small value of current in accordance with the following formula

$$PL = RI^2, \tag{1}$$

where R is the resistance and PL is power losses.

The current, which is produced by each layer, depends on the following factors:

1. Number of photons with energies higher than the band gap;
2. Absorption coefficients;
3. Thicknesses of the layers.

The number of absorbed photons can be controlled by adjusting the gap width of each layer, the absorption coefficient of the material also depends on the characteristics of the material constituting each layer. Consequently, the current of each layer can be controlled by varying the layer thickness. For example, if the number of photons entered in each layer is the same, the layers with low coefficient of absorption must be thicker than the other, to absorb the same number of photons as the other layers, as well as produce the same number of electron-hole pairs. MSC must be developed and designed for the gap width and thickness of each layer, taking into account the current produced by each layer separately.

Another way to increase the efficiency of solar cells is the use of such cells, each of which uses a certain part of the spectrum of solar radiation for the production of electric current. Tandem solar cells can be used as single or serial connection, where the current in both cases is similar. The structure of the solar cell in the series connection is very simple, but because of limitations caused by wide bandgap, solar cells of a single compound, from the viewpoint of efficiency, are more optimal.

The most common method of creating tandem solar cells is their growing even when the layers are built up sequentially on the substrate and provide a tunneling contact of layers in individual cells. Due to increasing of number of band gaps, the efficiency of the cell also icreases. The upper part of cell has the largest width of band gap; it absorbs photons which have higher energy of the spectrum of incident light, while the lower part of the cell has a small band gap width, and hence, provides the absorption of the low energy photons [25].

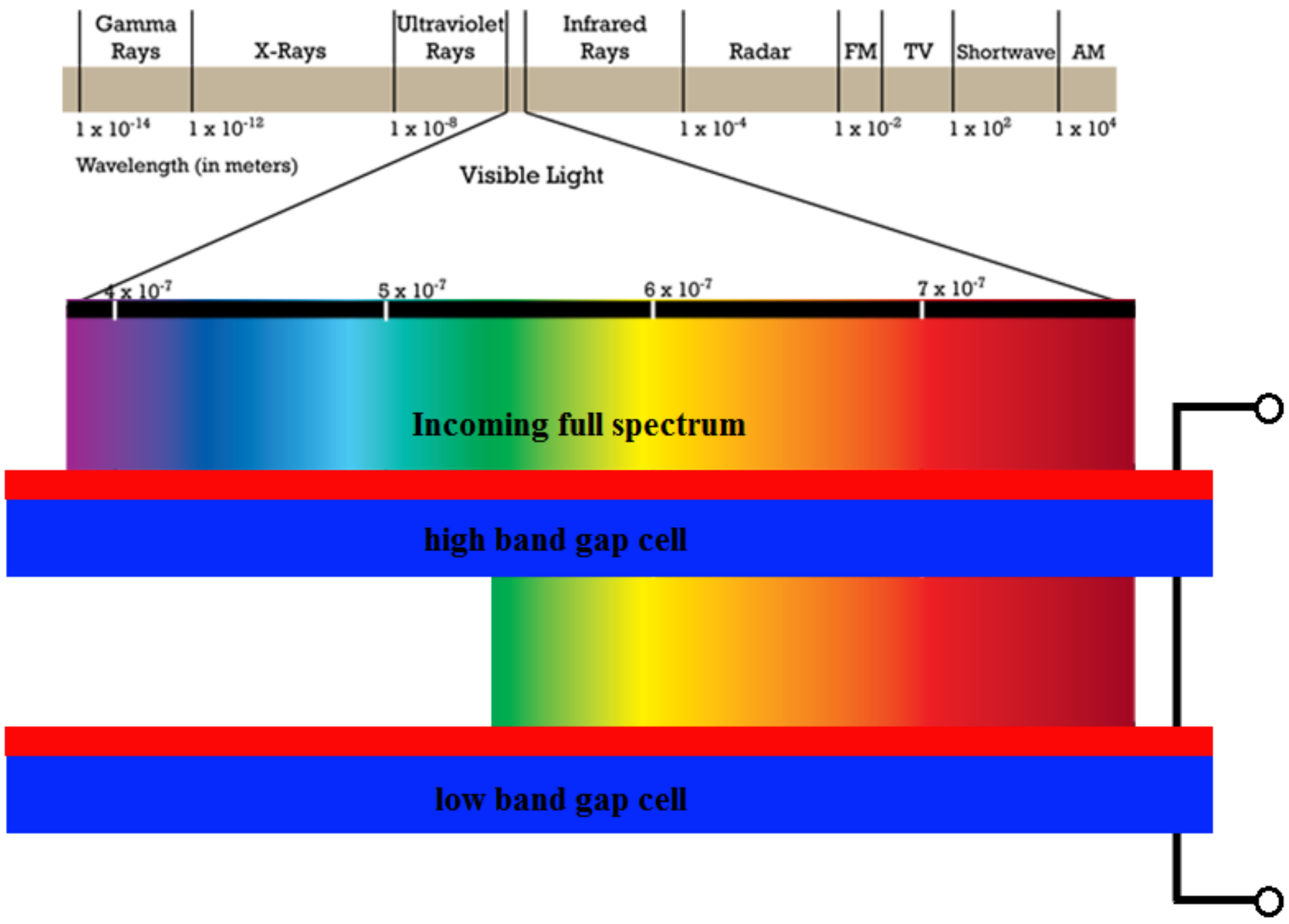

**Fig. 3.** Propagation of electromagnetic radiation, including ultraviolet, visible and infrared parts of the spectrum through the tandem solar cell

Range of the incident solar light on the upper part of solar cell includes all wavelengths, including ultraviolet, visible and infrared wavelengths. The absorption coefficient of the short-wave is high enough, so the bulk of the blue light is absorbed in the upper layer close to the surface, generating electric charge carriers. The first solar cell absorbs photons with the energy higher than the bandgap width and photons with lower energy than the width of bandgap pass through it and are absorbed in the next cell. The charge carriers generated by the short-wave part of the light spectrum, diffuse into the cell, collecting on the p-n-junction or recombine with the majority carriers in the bulk or on the junction. If all the charge carriers are collected near the contact, and do not recombine in any other part of the solar cell, the efficiency increases. This means that recombination, regardless of whether it happens, in the front or rear of the cell, directly influences the efficiency. In an initial approximation tandem multilayer solar cells similarly to single-junction cells are connected in series, so their open-circuit voltage is the sum of the basic cells voltages, and the short circuit current is equal to the minimum value of short circuit current of the basic cell. For this reason, any operation multilayer tandem solar cell can be obtained directly from the operations of the basic cells. The current density J is obtained by superposition of the diode current and the photo-generated current.

$$J = J_{ph} - J_{01}(\exp(qV/KT) - 1) - J_{02}(\exp(qV/2KT) - 1), \quad (2)$$

where $J_{ph}$ is photocurrent density; $J_{01}$ is saturation current density in perfect darkness; $J_{02}$ is current density in imperfect dark.

The density of the photocurrent and the dark current density obtained from the total light flux and the total density of the dark current generated in the emitter, base and the depletion region of the solar cell [26]

$$J_{ph} = J_{emitter} + J_{base} + J_{depleted} \quad (3)$$

$$J_{emitter} = \left[\alpha F(1-R)qL_p/(\alpha^2 L_p^2 - 1)\right]\times$$
$$\left[\frac{\frac{S_pL_p}{D_p} + \alpha L_p - \exp(-\alpha(d_e - W_n))\left(\frac{S_pL_p}{D_p}\cosh\frac{(d_e-W)}{L_p} + \sinh\frac{(d_e-W)}{L_p}\right)}{\left(\frac{S_pL_p}{D_p}\right)\sinh\left(\frac{(d_e-W)}{L_p}\right) + \cosh\left(\frac{(d_e-W)}{L_p}\right)} - \alpha L_p \exp(-\alpha(d_e - W_n))\right]$$

(4)

$$J_{base} = \left[qF(1-R)\alpha L_n/(\alpha^2 L_n^2 - 1)\exp\left((-\alpha(d_b - W_n + W))\right)\right]\times$$
$$\left[\alpha L_n - \frac{\frac{S_nL_n}{D_n}\left(\cosh\frac{(d_e-W_n)}{L_n}\right) - \exp(-\alpha(d_b - W_p)) + \sinh\frac{(d_b-W_p)}{L_n} + \alpha L_n \exp(-\alpha(d_b - W_p))}{\left(\frac{S_nL_n}{D_n}\right)\sinh\left[\frac{(d_b-W_p)}{L_n}\right] + \cosh\left[\frac{(d_b-W_p)}{L_n}\right]}\right]$$

(5)

$$J_{depleted} = qF(1-R)\exp(-\alpha(d_e - W_n))\left(1 - \exp(-\alpha W)\right)$$

(6)

$$J_{01} = J_{01,emitter} + J_{01,base}$$

(7)

$$J_{01,emitter} = q\frac{n_i^2}{N_D}\frac{D_p}{L_p}\left[\frac{\left(\frac{S_pL_p}{D_p}\cosh\frac{(d_e-W_n)}{L_p} + \sinh\frac{(d_e-W_n)}{L_p}\right)}{\frac{S_pL_p}{D_p}\sinh\left(\frac{(d_e-W_n)}{L_p}\right) + \cosh\left(\frac{(d_e-W_n)}{L_p}\right)}\right]$$

(8)

$$J_{01,base} = q\frac{n_i^2}{N_A}\frac{D_n}{L_n}\left[\frac{\left(\frac{S_n L_n}{D_n}\cosh\frac{(d_b - W_p)}{L_n} + \sinh\frac{(d_b - W_p)}{L_n}\right)}{\frac{S_n L_n}{D_n}\sinh\left(\frac{(d_b - W_p)}{L_n}\right) + \cosh\left(\frac{(d_b - W_p)}{L_n}\right)}\right] \quad (9)$$

$$J_{02} = \frac{Wn_i}{2(V_d - V)\tau} \quad (10)$$

where Q is the charge; F is the flow of incident photons; α is the coefficient of optical absorption; R is anti-glare reflection; $N_i$ is concentration of carriers; $N_A$ and $N_D$ are concentrations of acceptor and donor carriers, respectively; $d_e$ is the thickness of the emitter; $d_b$ is the thickness of the base; $L_p$ is the length of the penetration of hole in the emitter; $L_n$ is the length of penetration of the electrons into the base; $S_p$ is the surface recombination rate of holes in the emitter; $S_n$ is the surface recombination velocity of the electrons in the base; $D_p$ is the penetration rate of holes in the emitter; $D_n$ is the penetration rate of electrons in the base; τ is the lifetime of the charge carriers.

In what follows the charge transfer due to flux of incident photons in the basic unit is referred to as TF, the voltage on p-i-n- junction as $V_d$, the thickness of the depletion region of the emitter is $W_n$, the thickness of the depletion region on the basis is $W_p$ and the total thickness of the depletion region is W [27]

$$V_d = KT\log\left(\frac{N_D N_A}{n_i^2}\right) \quad (11)$$

$$W = \sqrt{2\varepsilon \frac{N_D + N_A}{N_D N_A}\left(V_d - V - 2KT\right)} \quad (12)$$

$$W_n = W / \left(1 + N_D / N_A\right) \quad (13)$$

$$W_p = W - W_n \quad (14)$$

where K is the Boltzmann constant, ε is the dielectric constant, and T is the temperature. The important thing is that there F and α depend on the wavelength and $D_p$, $D_n$, $L_P$, $L_N$, τ depend on the carrier concentration (impurity).

This section presents the research, aimed increasing the efficiency of solar cells and designing the structure of the solar cell with the most optimal efficiency, using single-junction solar cells whose efficiency is studied in the previous section. In order to do this we conducted a study of new multi-junction tandem a-SiC:H/a-SiH solar cells structures by use of the program AMPS-1D.

In the paper [28] experimental studies of the conductance properties of amorphous n-p-junctions in application to a tandem solar cells such as a-Si:H/HIT with a thin intermediate layer have been conducted. In particular, in this paper [28] the effect of the activation energy and the Urbach parameter of tunneling and recombination transitions have been studied, and it is shown that the best efficiency of the tandem cell is equal to 9.75%, which is achieved at an open circuit voltage $V_{oc}$ = 1430 mV, a current density $J_{oc}$ = 10.51 mA/cm$^2$ with fill factor FF = 0.65.

In order to increase the efficiency of the solar cell let us consider a multi-layer tandem solar cell which is composed of two a-SiC:H/a-Si:H cells with the parameters used in the simulation in the previous sections. These elements form the contact by the back sides, so that the order of the layers of the solar cell is depicted in Figure 5.

This solar cell is formed of two layers of a-SiC:H, a single layer of p-type and the other n-type layer in the top solar cell and two layers of a-Si:H, one of which is p-, and the other is n-type placed on the bottom of single-junction cell. Also, between these two cells the intermediate layer a-Si:H aiming to create the p - i- n – junction is placed.

As the upper protective TCO ($SnO_2$) layer of 200 nm thick is used to reduce the reflection, the lower layer of the solar cell is a reflection ZnO layer of 500 nm thick.

The parameters used in the numerical simulation of this solar cell using AMPS - 1D are shown in the Table 2.

The above numerical simulation allowed us to obtain current-voltage characteristics of the colar cell.

In this simulation, the thickness of the a-SiC:H p-layer is chosen equal to 100 nm, and the thickness of a-SiC:H n-layer is 200 nm, and these values were kept constant. The thickness of the a-Si:H p-layer is 1000nm, and the thickness of the a-Si:H n-layer is 3000nm. Thickness of i-layer in numerical simulation varied from 100 nm to 900 nm in order to obtain the maximum efficiency.

The Fig. 5 below depict the dependence of the main parameters of the solar cell, such as open circuit voltage $V_{oc}$, short circuit current density $J_{sc}$, fill factor FF and the efficiency on the thickness of a-Si:H i-layer. The results show that the highest efficiency equal to 22.6% of the studied solar cell is achieved at the thickness of 2700 nm under the standard sunlight AM 1.5 spectra.

This efficiency is higher than the efficiency of single-junction a-Si:H solar cell under the same conditions.

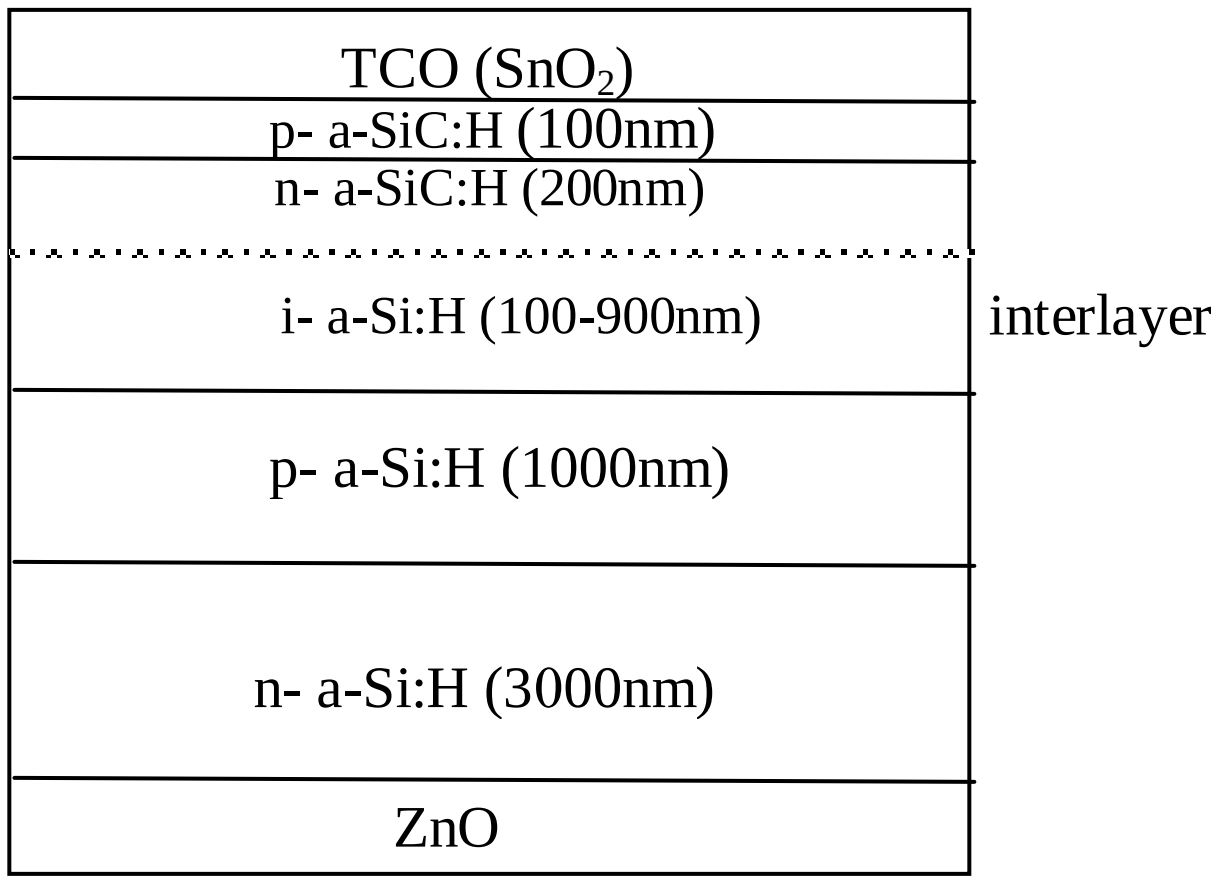

**Fig.4.** Schematic sketch of a-SiC:H/a-Si:H multi-junction tandem solar cell

**Table 2.** Parameters of tandem multijunction a-Si:H/a-SiC:H solar cell

| Material | Band gap (eV) | Conductivity type | Conduction Band | Valence Band | Electron Affinity (eV) | Electron Mobility (cm2 /v/s) | Hole Mobility (cm2 /v/s) | Free Carrier Concentration ($cm^{-3}$) | Relative Permittivity |
|---|---|---|---|---|---|---|---|---|---|
| **$SnO_2$:F** | 3.7 | P | $2.2*10^{18}$ | $1.8*10^{19}$ | 4.8 | 60.0 | 6.0 | $1.3*10^{19}$ | 9.0 |
| **a-SiC:H** | 1.9 | P | $2.5*10^{20}$ | $2.5*10^{20}$ | 4.0 | 20.0 | 2.0 | $3.0*10^{16}$ | 11.9 |
| **a-SiC:H** | 1.9 | N | $2.5*10^{20}$ | $2.5*10^{20}$ | 4.0 | 20.0 | 2.0 | $3.0*10^{18}$ | 11.9 |
| **a-Si:H** | 1.75 | N/i | $2.5*10^{20}$ | $2.5*10^{20}$ | 3.8 | 20.0 | 2.0 | $8.0*10^{19}$ | 11.9 |
| **a-Si:H** | 1.75 | P | $2.5*10^{20}$ | $2.5*10^{20}$ | 3.8 | 20.0 | 2.0 | $8.0*10^{17}$ | 11.9 |
| **a-Si:H** | 1.75 | N | $2.5*10^{20}$ | $2.5*10^{20}$ | 3.8 | 20.0 | 2.0 | $8.0*10^{19}$ | 11.9 |
| **ZnO:B** | 3.3 | N | $2.2*10^{18}$ | $1.8*10^{19}$ | 4.5 | 33.0 | 8.0 | $8.0*10^{18}$ | 9 |

**Fig.5.** Dependence of output parameters of a-SiC:H/a-Si:H multijunction tandem solar cell on the i-interlayer thickness

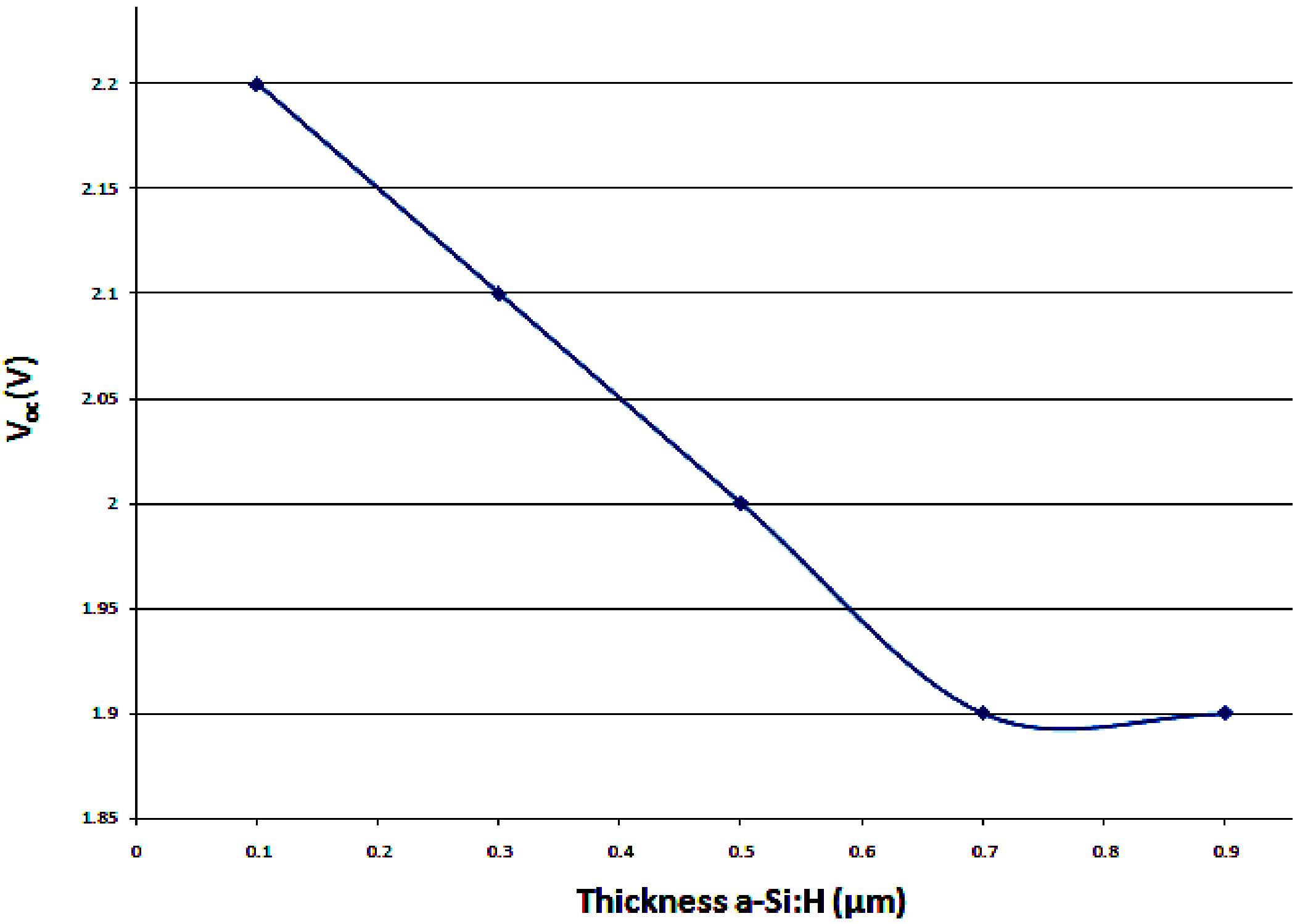

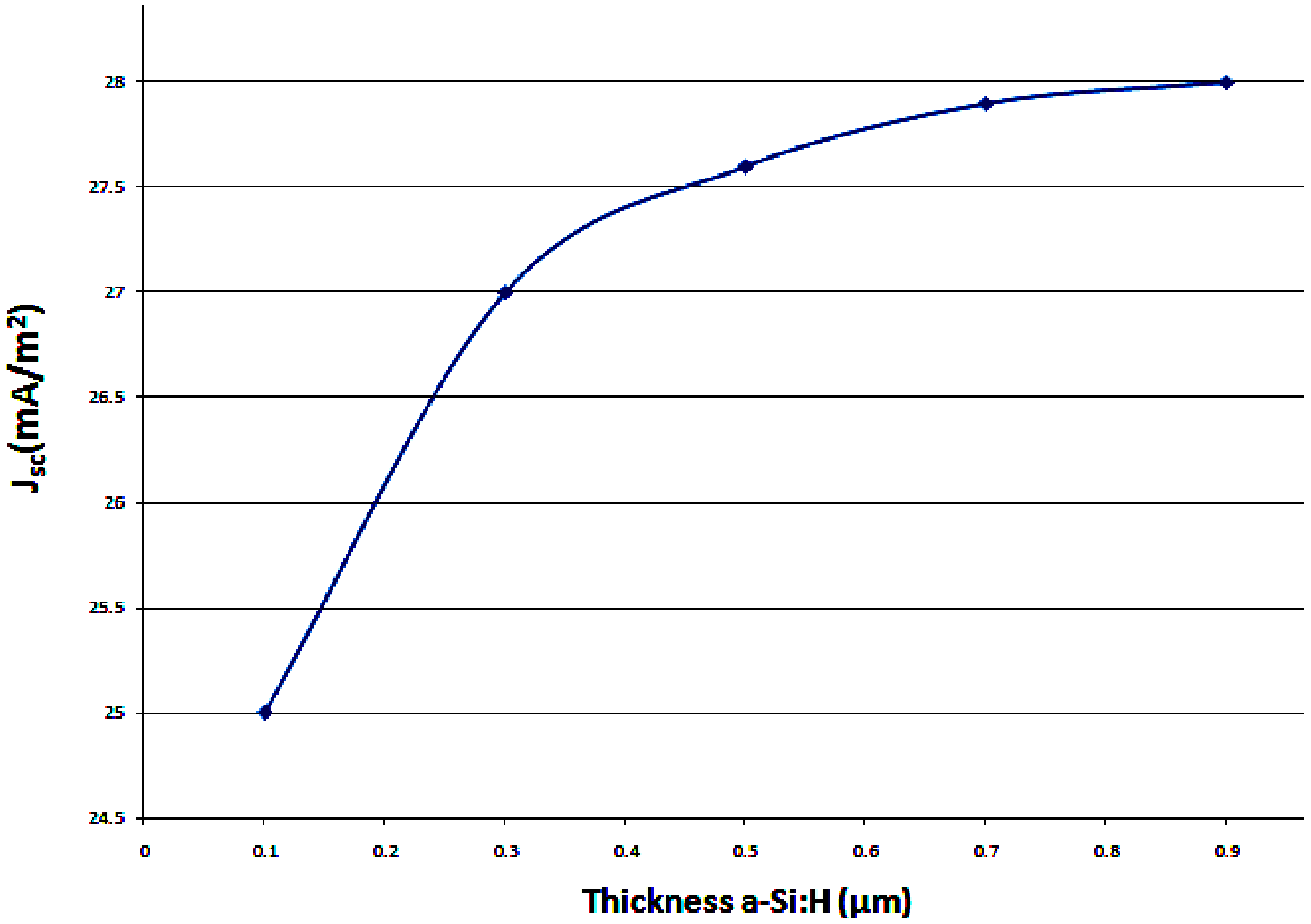

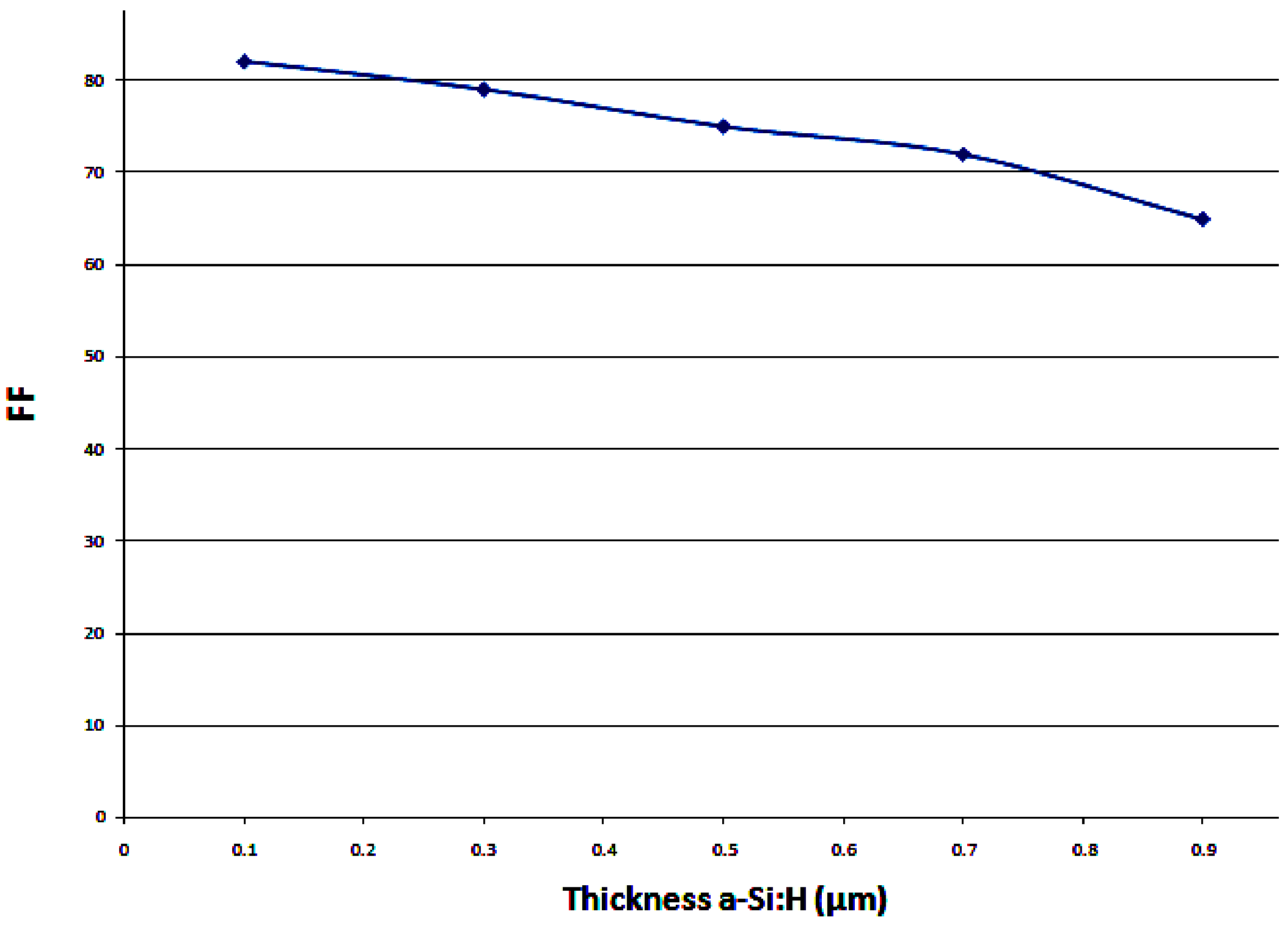

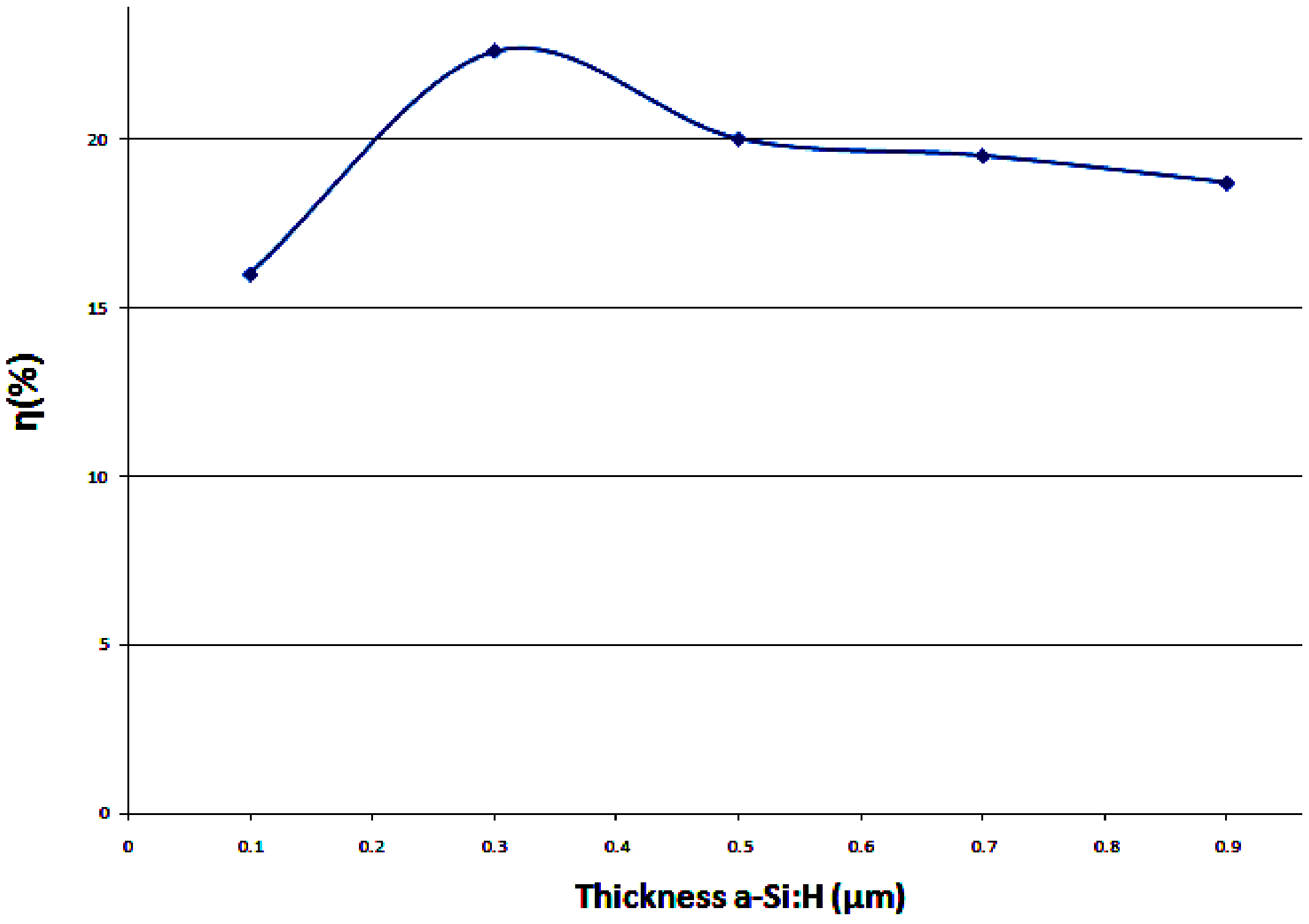

## 4. Conclusion

In this paper we carried out theoretical study of the general issues of the efficiency of single-junction and multi-junction tandem solar cells. Implementation of numerical simulations by the use of AMPS-1D program of one-dimensional analysis of microelectronic and photonic structures for the analysis of hydrogenated silicon solar cells allowed us to formulate the optimal design of new kind of multi-junction tandem solar cells, providing its most efficient operation.

The numerical analysis of a-Si:H/a-SiC:H single-junction solar cell whith doped i-layer used as the intermediate absorbing layer (a-Si:H) placed between layers of p-type (a-SiC:H) and n-type (a-Si:H) has been conducted. It is established that, after optimizing the solar cell parameters, its highest efficiency of 19.62% is achieved at 500 nm thickness of i-layer. The maximum experimental value of the similar solar cell efficiency is 10.1% (Kabir et al. [4]).

The optimization of the newly developed multi-junction structure of a-SiC:H/a-Si:H tandem solar cell has been conducted. It has been shown numerically that its highest efficiency of 22.6% is achieved at the thickness of 270 nm of intermediate i-layer.